# A NOVEL TECHNIQUE TO DETERMINE ATMOSPHERIC ION MOBILITY SPECTRA

K.L. Aplin

Space Science and Technology Department, Rutherford Appleton Laboratory, Chilton, Didcot, Oxon, OX11 0QX, UK

ABSTRACT: Detailed tropospheric ion measurements are needed to improve understanding of the electrical microphysics affecting clouds. Additionally, atmospheric ion mobility spectra can be used to identify ion growth processes leading to condensation nucleus formation. However these measurements are rare, particularly in the troposphere where the majority of clouds form. Developments in the operating theory of the classical instrument for ion measurement, the aspirated cylindrical capacitor, are described, which enable ion mobility spectrum information to be extracted from the rate of voltage decay of the aspirated capacitor in air. In this paper, data from historical balloon-borne ion counter ascents will be reanalysed to extract new ion mobility spectra from simple voltage time series. Such data recovery will increase the amount of atmospheric ion spectra available for analysis.

## INTRODUCTION

Atmospheric ions are one source of atmospheric aerosol particles. If, as has been proposed, particles formed from ions can influence cloud formation under some conditions [*Yu and Turco*, 2001], atmospheric ion measurements may reveal unique data on atmospheric electrical influences on clouds and climate. Although long-term observations of ion spectra exist at the surface [*Hõrrak et al*, 2000], data is very limited in cloud-forming regions. Free tropospheric ion spectra usually require a dedicated mass spectrometer, flown on a plane [*Arnold et al,* 1998].

The ventilated cylindrical capacitor or *Gerdien condenser* [*Gerdien,* 1905] is a long-established instrument for atmospheric ion measurements. It consists of a cylindrical outer electrode containing a coaxial central electrode, with a fan to draw air through the electrodes. With an appropriate bias voltage applied across the electrodes, a current flows in proportion to the ion concentration. The conductivity of the air can be inferred from the rate of decay of voltage across the electrodes. For surface measurements, this technique has generally been superseded by direct measurements of the current at the central electrode. Contemporary instruments based on this principle deploy modern electronics and computer control to use both the Current Measurement and Voltage Decay modes for self-calibration [*Aplin and Harrison*, 2000; 2001]. However the Voltage Decay mode still remains preferable for balloon-borne atmospheric ion instrumentation due to its simplicity in operation, measurement and telemetry.

Surface measurements with modern instrumentation showed that although generally comparable, the two modes were not always completely consistent [*Aplin and Harrison*, 2001]. This motivated reconsideration of the classical principles of the Voltage Decay mode. The modified approach enables ion mobility spectra to be retrieved from voltage decay measurements. In this paper, the new technique is applied to calculate ion mobility spectra from historic conductivity-sonde data.

## CLASSICAL THEORY

Atmospheric conductivity $\sigma$ is related to both the atmospheric ion concentration *n*, and the average mobility $\mu$ of the air ion population. Molecular ions with $\mu > 0.5$ cm$^2$V$^{-1}$s$^{-1}$ are defined as "small ions" [*Hõrrak et al*, 2000]. $\sigma$ can be written as

$$\sigma = e \int_{0.5cm^2V^{-1}s^{-1}}^{\infty} n \quad d\mu \qquad \text{Eq 1}$$

where *e* is the charge on the electron. Eq 1 is commonly simplified to

$$\sigma_{\pm} = en_{\pm}\mu_{\pm} . \qquad \text{Eq 2}$$

Typical surface $\sigma$~5-100fSm$^{-1}$ depending on aerosol pollution levels. Atmospheric ionisation is caused primarily by cosmic rays, though near the surface local radioactivity dominates. The increase in cosmic ray ionisation and ion mean free path with decreasing atmospheric pressure cause $n$, $\mu$ and $\sigma$ to rise with height. There is also a latitudinal and solar cycle variation in $\sigma$ due to the varying penetration of cosmic rays into the atmosphere.

For a classical Gerdien condenser in Current Measurement mode, if the ions reaching the central electrode are constantly replenished, $\sigma$ is proportional to the ion current measured at the central electrode. (The "conductivity measurement régime" requires an adequate ventilation speed and bias voltage, and can be verified by a linear response of measured current to a changing bias voltage). Using Eq 1, and considering the charge arriving at the central electrode per unit time results in the routinely-used equation for calculating $\sigma$ from current measurements with a Gerdien condenser, considered as a classical capacitor with capacitance $C$ [*e.g. MacGorman and Rust,* 1998].

$$\sigma_{\pm} = \frac{\varepsilon_0 i_{\pm}}{C V_{\pm}} . \qquad \text{Eq 3}$$

In the Voltage Decay mode, a voltage across the electrodes will decay due to the current $i$ flowing through the air, which has a large resistance $R$. If the charge on the capacitor is $Q$, elementary circuit analysis gives

$$\frac{dQ}{dt} R = -\frac{Q}{C} \qquad \text{Eq 4}$$

from which the familiar expression for the exponential decay of charge from a capacitor is derived. From this, the $\sigma$ measured by a Gerdien condenser in Voltage Decay mode is inversely related to the time constant $\tau$ of the decay

$$\sigma = \frac{\varepsilon_0}{\tau} . \qquad \text{Eq 5}$$

Eq 5 has been the standard expression for calculating $\sigma$ from voltage decay measurements throughout the history of the Gerdien condenser instrument, using $\tau$ determined from a time series of voltage measurements [*e.g. Swann,* 1914; *Aplin and Harrison*, 2000]. Eq 4 assumes that the resistance of air is constant, to give an exponential rate of decay of charge from a Gerdien condenser. However, measurements show that a non-exponential decay rate is commonly observed [*Venkiteshwaran*, 1958; *Aplin*, 2000], suggesting this approximation may not be universally appropriate.

## MODIFICATION TO CLASSICAL MOBILITY SPECTRUM THEORY

The minimum mobility of ion contributing to $\sigma$ measurement, the *critical mobility* $\mu_c$, [*e.g. MacGorman and Rust, 1998*] is usually approximated from the ventilation speed through a Gerdien condenser $u$, its length $L$ and the bias voltage $V$:

$$\mu_c = \frac{ku}{V} . \qquad \text{Eq 6}$$

where $k$ is a constant related to the dimensions of the Gerdien condenser [*e.g. MacGorman and Rust*, 1998]. The functional dependence of $\mu_c$ on the bias voltage can be exploited to compute ion mobility spectra from the changing ion current at the central electrode [*e.g. Dhanorkar and Kamra*, 1993].

During a Voltage Decay measurement the voltage across the electrodes decreases; from Eq 6, $\mu_c$ also varies during the decay. The number of ions contributing to the measurement therefore changes as a function of the decay voltage. This is identical to voltage decay through a voltage-dependent resistor *R(V)* in a parallel RC circuit, and Eq 4 can be rewritten to include this,

$$\frac{dq}{dt} R(V) = -\frac{Q}{C} . \qquad \text{Eq 7}$$

Solving Eq 7 gives a modified form of the classical decay, in which $V_t$ is implicitly defined:

$$V_t = V_0 \exp\left( \frac{-1}{C} \int_0^t \frac{dt}{R(V_t)} \right) \qquad \text{Eq 8}$$

*R(V)* and the shape of the decay curve are related to the shape of the ion mobility spectrum. If all ions have the same mobility (referred to here as a *flat* ion mobility spectrum) *R(V)* is constant during the decay and Eq 8 reduces to Eq 5, the classical exponential decay. For *all other* spectra, (here called *variable* ion spectra), *R(V)* is not constant and there is no theoretical basis on which expect an exponential decay. This is the general case in the atmosphere [*Hõrrak et al*, 2000].

The voltage decays expected from prescribed atmospheric ion mobility spectra can be calculated by numerical integration of Eq 8. The simplest ion spectrum is flat (Figure 1a). As the number of ions in each mobility category is constant, the effective resistance does not change during the decay, and the voltage decay is exponential (Figure 1b). Another simple shape of mobility spectrum is based on existing average ion spectra [*Hõrrak et al*, 2000]. (Figure 1c). The voltage decay from this spectrum (Figure 1d) is not exactly exponential, with only 73% of its variance explained by an exponential model. A typical voltage decay time series in atmospheric air would only yield an exponential decay to a first approximation. The difference from the exponential decay is related to the shape of the spectrum; therefore, voltage decays contain spectral information, which can be obtained using a suitable inversion technique [*Aplin,* 2003].

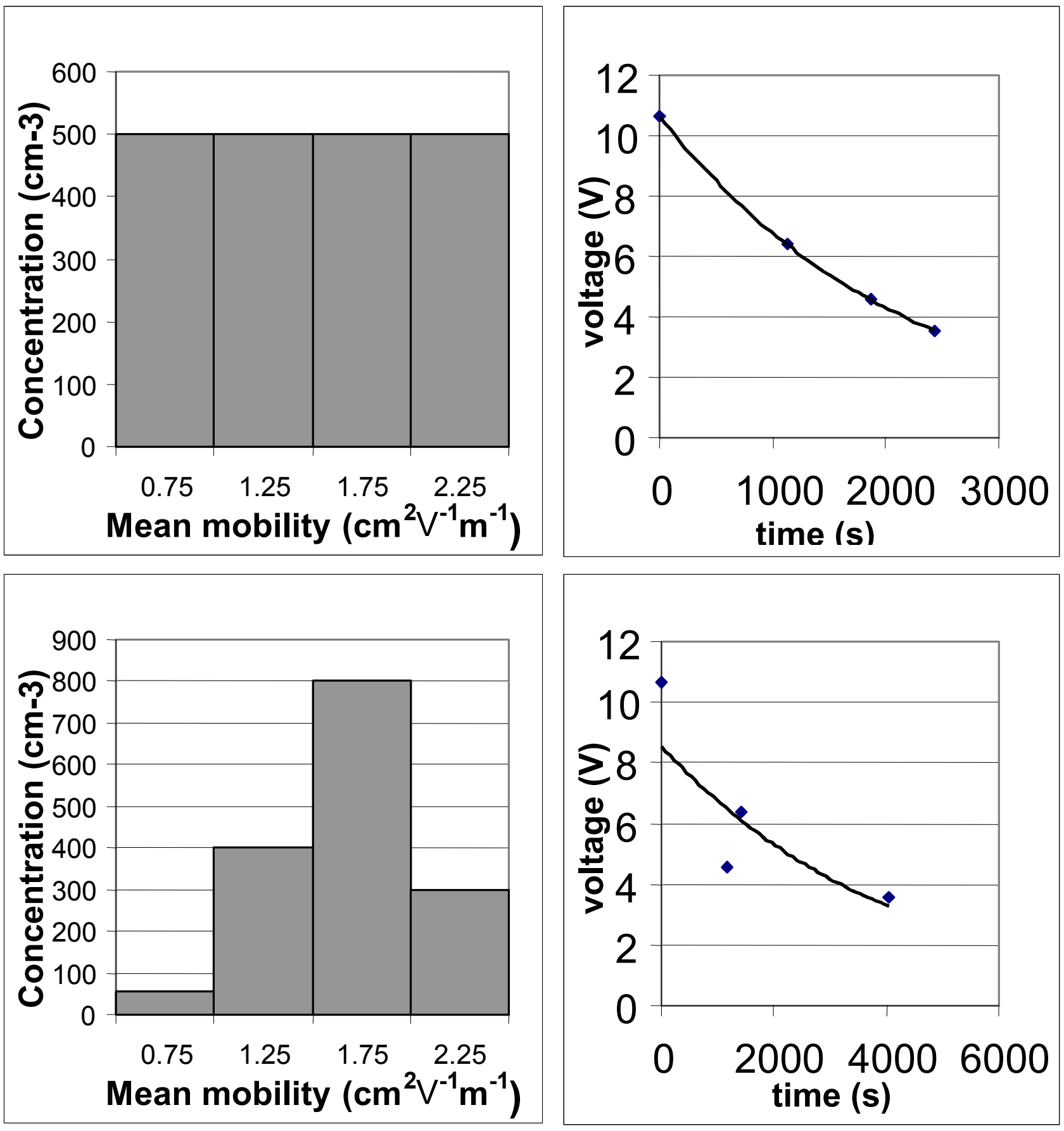

Figure 1 Sample ion mobility spectra with equivalent voltage decays. Ion mobility spectra (left) are simplified and replaced by four equal mobility bands, with the average mobilities indicated. Concentrations and mobilities are typical for surface atmospheric levels. Calculated voltage decays (right) assume a Gerdien condenser with dimensions given in [*Aplin and Harrison,* 2000]. a) A flat ion spectrum inverts to an exponential decay, c). b) a Gaussian ion spectrum inverts to a non-exponential decay, d).

FREE TROPOSPHERIC ION SPECTRA FROM VOLTAGE DECAY MEASUREMENTS

At the surface, the effect of the ion mobility spectrum on the shape of the voltage decay may be difficult to distinguish from other sources of variability. Effects such as turbulence and deposition of charged aerosol particles on the electrodes can contribute towards a non-exponential decay [*Aplin,* 2000]. In the free troposphere, where aerosol concentrations are minimal, non-exponential decays have also been observed [*Venkiteshwaran,* 1958] for which there are fewer obvious explanations. Free tropospheric voltage decay measurements have therefore been selected for inversion here, as the deviations of the decays from the exponential are more likely to be due to true spectral variations.

Data from historic conductivity-sonde ascents is usually presented as conductivity as a function of altitude (Eq 5), with the "raw" voltage decays rarely shown. However Venkiteshwaran [1958] showed the raw decays, suitable for inversion. A modified radiosonde carrying a Gerdien condenser measured three decays at 200, 250 and 300 mbar on 3 February 1958 (corresponding to a mid- to upper troposphere altitude of 9-10km) over Pune, India. The critical mobility of the instrument was estimated from its dimensions, and a flow rate of $3ms^{-1}$ was assumed. To calculate the spectrum reduced mobility was used, as is customary, to permit comparison of ion species at different heights [*e.g. MacGorman and Rust*, 1998].

The ion spectrum was was calculated by a numerical solution of Eq 8 applied to the digitised voltage decay data. The total ion concentration $n$ was estimated by integrating the area under the ion mobility spectrum to be $n$~3900 $cm^{-3}$, increasing with height, as would be expected. The absolute value of $n$ can be verified if the ionisation rate $q$ is known, as the free tropospheric ion concentration $n$ is limited solely by self-recombination with coefficient $\alpha$

$$n = \sqrt{\frac{q}{\alpha}} \qquad \text{Eq 9}$$

Gringel *et al* [1986] give $q$ at a geomagnetic latitude (GL) of 36 °N in 1958 as $20cm^{-3}s^{-1}$ at 10km. As the air over Pune experienced similar energy cosmic ray penetration (GL ~ 30°N) then it is legitimate to estimate $n$ from Eq

9. $n \sim 4000$ cm$^{-3}$, indicating that the integrated value of $n$ is roughly correct. The reduced mobility was normalised using typical $\sigma$ [Gringel *et al,* 1986] and the calculated $n$, to reduce the error from the estimates made in the mobility calculation. The calculated ion spectrum is shown in Figure 2.

Unfortunately no tropospheric ion mobility spectra have yet been found in the literature for comparison, but this spectrum shows qualitative agreement with surface and stratospheric ion spectra. For example, the 300mbar spectrum has a peak in the 1-2 cm$^2$V$^{-1}$s$^{-1}$ range, as do the ion spectrum measurements made by *Hõrrak et al* [2000] and *Gringel et al* [1986]. Venkiteshwaran [1958] suspected the radiosonde passed through 5/8 cirrus cloud at 250 mbar, which is consistent with the peak in larger (lower mobility) ions unique to the 250mbar spectrum shown in Figure 2.

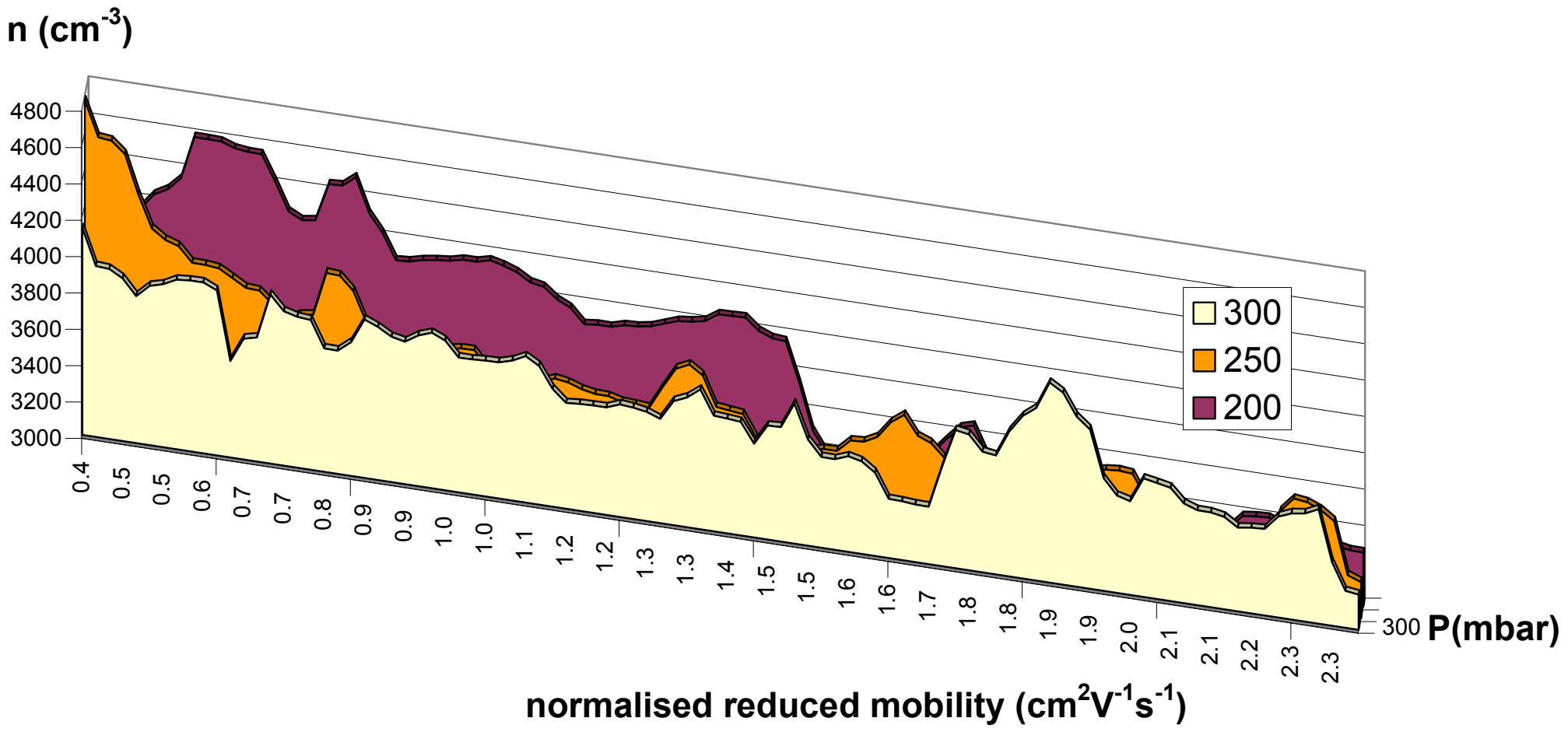

Figure 2 Smoothed reduced mobility spectrum for tropospheric ions at three pressures over Pune, India, 3 Feb 1958

## CONCLUSION

The physical principles of the Gerdien condenser have been reconsidered, and the classical theory modified to take account of the effect of ion spectrum variations. This has permitted inversion of historic $\sigma$ data. Three tropospheric ion spectra have been calculated which are consistent with other measured ion mobility spectra. Further historic (and contemporary, if available) voltage decay data are clearly needed for corroboration. An atmospheric measurement campaign would be desirable to further establish this method, and, ultimately, determine the role of atmospheric small ions in cloud condensation nucleii formation.